\documentstyle[12pt,epsfig]{article}
\begin{document}
\baselineskip=23pt

\begin{center}
{\Large\bf AdS Dynamics for Massive Scalar Field:\\[0pt]
 exact solutions vs. bulk-boundary propagator}

\bigskip
\bigskip

Zhe Chang\footnote{Email: changz@hptc5.ihep.ac.cn.}  and
Cheng-Bo Guan\footnote{Email: guancb@hptc5.ihep.ac.cn.}

{\em
Institute of High Energy Physics, Academia Sinica\\
P.O.Box 918(4), Beijing 100039, China\\}

\medskip

Han-Ying Guo\footnote{Email: hyguo@itp.ac.cn.}

{\em 
Institute of Theoretical Physics, Academia Sinica\\
P.O.Box 2735, Beijing 100080, China}

\end{center}

\bigskip
\bigskip

$AdS$ dynamics for massive scalar field is studied both by solving exactly 
the equation of motion and by constructing bulk-boundary propagator.
A Robertson-Walker-like metric is deduced from the familiar $SO(2,n)$
invariant metric. The metric allows us to present a time-like Killing 
vector, which is not only invariant under space-like transformations but 
also invariant under the isometric transformations of $SO(2,n)$ in certain
sense. A horizon appears in this coordinate system. Singularities of field 
variables at boundary are demonstrated explicitly. It is shown that there 
is a one-to-one correspondence among  the exact solutions and the bulk fields 
obtained by using the bulk-boundary propagator.
{\em keywords}: $AdS$ space; Scalar field; Exact solutions; Propagator; 
                Singularity
\newpage

\section{Introduction}

$\quad$ The interest on dynamics in Anti-de Sitter ($AdS$) space\cite{001} 
has been revived by a conjectured duality between string theory in the bulk of 
$AdS$ and conformally invariant field theory ($CFT$) living on the boundary
of $AdS$ \cite{002}. The $AdS/CFT$ correspondence gives an explicit relation
\cite{003,004} between Yang-Mills theory and string theory\cite{03a,04a}. 
It is wished that this duality will offer us a theory of strong coupling 
mysteries of $QCD$. Indeed, many interesting results\cite{005}--\cite{010}, 
including confinement, $\theta$ vacuum, glueball mass spectrum and topological 
susceptibility etc., have been obtained by simple calculations of supergravity.
This is qualitatively similar to what happens in strong coupling lattice $QCD$. 
Furthermore and curiously, the ratios of the glueball excited state masses and 
the lowest state mass\cite{006}--\cite{009} are in reasonable good agreement 
with the lattice computations. The $AdS/CFT$ correspondence states that the 
string theory on $AdS_d\times M_{D-d}$ is dual to a conformal field theory 
living on the spacetime boundary. To each field $\Phi_i$ there is a 
corresponding local operator ${\cal O}^i$ in the conformal field theory. 
A strong support for the conjectured duality comes from comparing spectra 
of the Type IIB string theory on the background of $AdS_5\times S^5$ and 
low-order correlation functions of the $3+1$ dimensional ${\cal N}=4$ 
$SU(N)$ super Yang-Mills theory.

In order to show explicitly the $AdS/CFT$ duality or to investigate its 
delicious implications, dynamics in $AdS$ should be studied extensively in 
various directions. In this paper, we study $AdS$ dynamics for massive scalar 
field both by solving exactly the equation of motion and by constructing 
bulk-boundary propagator. A Robertson-Walker-like metric is deduced from the 
familiar $SO(2,n)$ invariant metric in an $(n+2)$-dimensional embedding space.
Thus, we can present a time-like Killing vector in this metric. However, it is
not only invariant under space-like transformations but also invariant under
the $AdS$ transformations of $SO(2,n)$ in certain sense. A horizon appears in
this coordinate system. We obtain a bulk-boundary propagator for $AdS$ with 
correct signature. Singularities of wavefunctions at the boundary are also
demonstrated explicitly. It is shown that there is a one-to-one correspondence
among  the exact solutions and the bulk fields obtained by using of the 
bulk-boundary propagator.

The paper is organized as follows. In Section 2, global geometric properties
of $AdS$ are discussed. Relations between the explicit $SO(2,n)$ invariant 
metric and the Hua's metric are shown. In Section 3, we introduce a coordinate
of $AdS$ with a Robertson-Walker-like metric and a space-transformation 
invariant time-like Killing vector. The Penrose diagram is drawn for this 
coordinate and shows that there is a horizon. $AdS$ is divided by the horizon 
into two separate parts, {\em i.e.}, region I and region II. Physics in 
different regions is discussed, separately. We obtain exact solutions of free 
massive scalar field on region I by making use of the variable-separating
method in Section 4. Discrete mass spectrum is got by natural boundary
conditions on the horizon. Bulk-boundary propagator in the region I is 
discussed in section 5. Results obtained by solving exactly the equation of
motion and ones got by constructing the bulk-boundary propagator are in good 
agreement. Section 6 is devoted to the exact solutions of the massive scalar 
field in the region II. In Section 7, we present a bulk-boundary propagator 
in the region II.

\section{Some global geometric properties of $AdS$}

$\quad$ In an $(n+2)$-dimensional embedding space, the $(n+1)$-dimensional 
$AdS$ can be written as

\begin{equation} \label{sphere}
 \xi^0\xi^0-\sum_{i=1}^n\xi^i\xi^i+\xi^{n+1}\xi^{n+1}=1~,
\end{equation}
where the cosmological radial constant is normalized as one. It is easy to 
check that the isometric symmetry of $AdS_{n+1}$ is $SO(2,n)$.
It is well known that, at least, two charts of coordinates ${\cal U}_1$
($\xi^{n+1}\not= 0$) and ${\cal U}_0$ ($\xi^0\not= 0$) should be needed to 
describe $AdS$ globally.

In the chart $AdS_{n+1}\cap {\cal U}_1$, we introduce a coordinate system

\begin{equation}
 x^i=\frac{\xi^i}{\xi^{n+1}}~,~~~~~~(i=0,~1,~2,~\cdots,~n;~~\xi^{n+1}\not=0)~,
\end{equation}
for both parts ${\cal U}^+_1(\xi^{n+1}>0)$ and ${\cal U}^-_1(\xi^{n+1}<0)$~.\\
The $AdS_{n+1}\cap {\cal U}_1$, in the coordinate $(x^i)$, is described by

\begin{equation}
\begin{array}{l}
AdS_{n+1}\cap {\cal U}_1:~~~~~~\sigma(x^i,x^j)>0~,\\[0.5cm]
\sigma(x^i,x^j)\equiv 1+\displaystyle\sum_{i,j=1}^n\eta_{ij}x^i x^j ~,
~~~~~\eta={\rm diag}(1,~\underbrace{-1,~-1,~\cdots,~-1}\limits_{n})~.
\end{array}
\end{equation}
In the chart $AdS_{n+1}\cap {\cal U}_0$, we can also introduce a coordinate
system

\begin{equation}
\begin{array}{l}
\displaystyle y^0=\frac{\xi^{n+1}}{\xi^0}~,~~~~~~(\xi^0\not= 0)~,\\[0.5cm]
\displaystyle y^i=\frac{\xi^{i}}{\xi^{0}}~,~~~~~~(i=1,~2,~\cdots,~n;~\xi^0
\not=0)~,
\end{array}
\end{equation}
for both parts ${\cal U}^+_0(\xi^0>0)$ and ${\cal U}^-_0(\xi^0<0)$~.\\
In the coordinate $(y^i)$, the $AdS_{n+1}\cap {\cal U}_0$ is described as

\begin{equation}
AdS_{n+1}\cap {\cal U}_0:~~~~~~\sigma(y^i,y^j)>0~.
\end{equation}
The domains  $\sigma(x^i,x^j)>0$ and $\sigma(y^i,y^j)>0$~ are classical
domains.
At the overlap region $AdS_{n+1}\cap{\cal U}_0\cap {\cal U}_1$ of the two
charts ${\cal U}_1$ and ${\cal U}_0$, we have relations between the coordinates
$(x^i)$ and $(y^i)$

\begin{equation}
y^0=\frac{1}{x^0}~,~~~~~~y^i=\frac{x^i}{x^0}~.
\end{equation}
This shows clearly a differential structure of $AdS$.

The boundary $\overline{M}^n$ of the slice $AdS_{n+1}\cap{\cal U}^+_1({\rm or}~
{\cal U}^-_1)$ of $AdS$ is as

\begin{equation}
\overline{M}^n:~~1+\eta_{jk}x^jx^k=0~.
\end{equation}
This consists of the conformal space of Dirac\cite{001}. In Fig.1, we show a 
plot of $AdS$ and its boundary in the coordinate $(x^i)$.

\begin{figure}
\epsfxsize=90mm
\epsfysize=90mm
\centerline{\epsffile{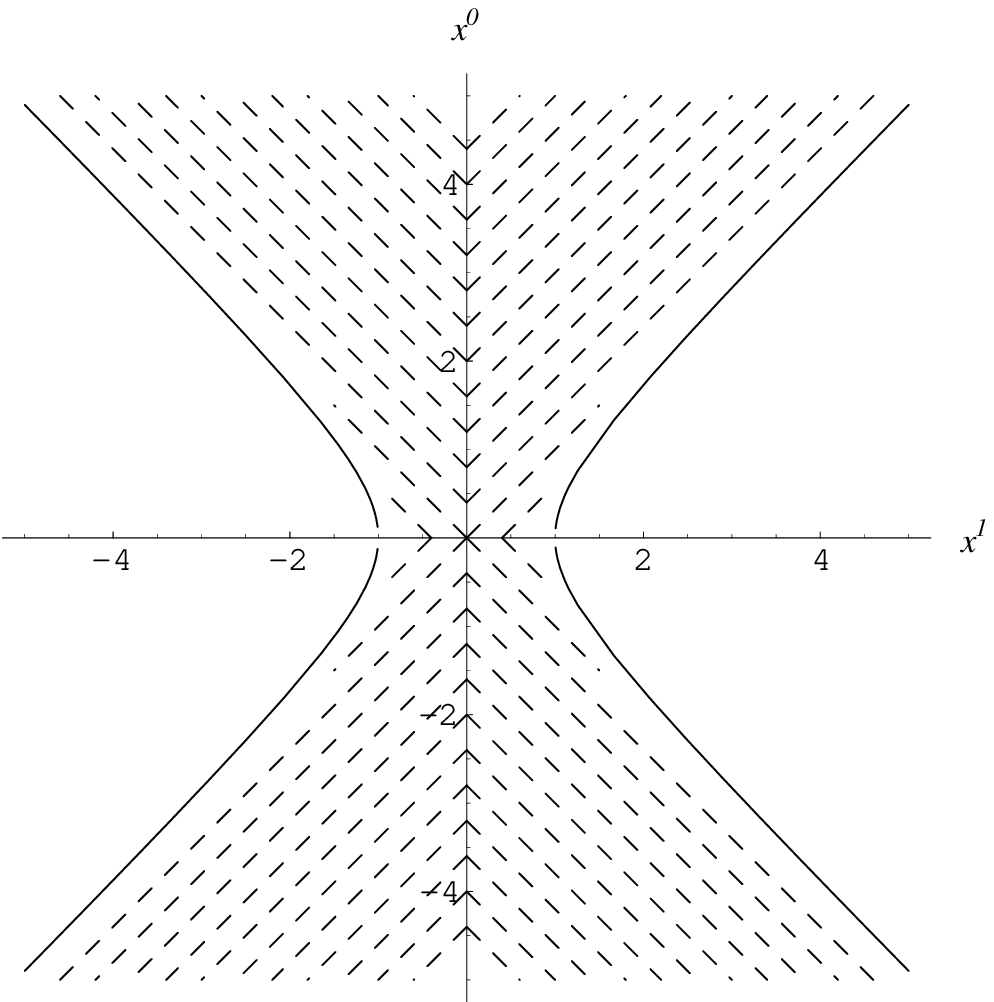}}
\vspace*{0.3cm}
\centerline{Fig.1~~$AdS_2$ in coordinate $(x^i)$ as classical
domain covered by dashing lines.}
\end{figure}
For the pseudo-sphere (\ref{sphere}), an explicit $SO(2,n)$ invariant metric
can be introduced as

\begin{equation}
 ds^2= \eta_{ij}d\xi^id\xi^j+d\xi^{n+1}d\xi^{n+1}~.
\end{equation}
In the coordinate $(x^i)$, the metric is reduced as

\begin{equation}
\begin{array}{rcl}
ds^2&=&\left(\sigma^{-1}\eta_{ij}-\sigma^{-2}\eta_{ik}\eta_{jl}x^kx^l
            \right)dx^idx^j \\[0.4cm]
   &=&-\displaystyle\frac{dxJ(I-x^{\prime}xJ)^{-1}dx^{\prime}}{1-xJx^{\prime}}~,
\end{array}
\end{equation}
where we have used the notations
$J\equiv{\rm diag}(-1,~\underbrace{1,~1,~\cdots,~1}\limits_{n})$, $x\equiv
(x^0,~x^1,~\cdots,~x^n)$~, and $x'$ the transport of the vector $x$. This is
obviously a generalization of the Hua's metric \cite{10a,10b} for manifolds
with Lorentzian signature.
It should be noticed that the transformations among the coordinate 
variables $(x^\alpha)~(\alpha=1,~2,~\cdots,~n)$ form a group ${\cal D}
\subset SO(1,n)$, which is the isometric transformations in $(\xi^1,~\xi^2,
~\cdots,~\xi^{n+1})$. Thus, we can refer to these transformations as 
space-like ones. The $x^0$ element of $(x^i)$ is not invariant under the 
so-called space-like transformations ${\cal D}$. We can not got a time-like 
Killing vector using $x^0$ only. To draw explicitly time-like geodesics, 
one should investigate space-like transformation invariant variables. 
In fact, under the transformations ${\cal D}$,
$\xi^0(\equiv\sigma^{-\frac{1}{2}}(x^i,x^j)x^0)$ is invariant.  Therefore, 
it is convenient to use the coordinate $(\xi^0,~x^1,~\cdots,~x^n)$ \cite{10c}.

\section{Spacetime geometry in $(\protect\xi^0,{\bf x})$ coordinate system}

$\quad$ By making use of the relations between the coordinates $(x^i)$ and
$(\xi^0,~x^\alpha)$

\begin{equation}
\begin{array}{l}
 \displaystyle \sigma(x^i,x^j)=\frac{1-{\bf x}{\bf x}^{\prime}}
  {1-\xi^0\xi^0}~,\\[0.5cm]
 \displaystyle x^0x^0=\frac{\xi^0\xi^0}{1-\xi^0\xi^0}(1-{\bf x}{\bf x}
  ^{\prime})~,
\end{array}
\end{equation}
we get a deduced Robertson-Walker-like metric in terms of the coordinate
$(\xi^0,~x^\alpha)$

\begin{equation}\label{eq5}
 ds^2=\frac{1}{1-\xi^0\xi^0}d\xi^0d\xi^0-(1-\xi^0\xi^0)
 \frac{d{\bf x}(I-{\bf x}^{\prime}{\bf x})^{-1}d{\bf x}^{\prime}}
      {1-{\bf x}{\bf x}^{\prime}}~,
\end{equation}
where the vector ${\bf x}$ denotes $(x^1,~x^2,~\cdots,~x^n)$ and ${\bf x}
^{\prime}$ the transport of the vector ${\bf x}$. \\
In the spherical coordinate $(x^1,~x^2,~\cdots,~x^n)\longrightarrow
(\rho,~\theta_1,~\cdots,~\theta_{n-1})$, the $SO(2,n)$ invariant
metric (\ref{eq5}) is of the form

\begin{equation}
 ds^2=\frac{1}{1-\xi^0\xi^0}d\xi^0d\xi^0
      -(1-\xi^0\xi^0)\left[(1-\rho^2)^{-2}d\rho^2
      +(1-\rho^2)^{-1}\rho^2d{\bf u}d{\bf u}^{\prime}\right] ~,
\end{equation}
where
 $$\begin{array}{rl}
    {\bf u}=&(\cos\theta_1,~\sin\theta_1\cos\theta_2,
              ~\sin\theta_1\sin\theta_2\cos\theta_3,
              ~\cdots,
              ~\sin\theta_1\cdots\sin\theta_{n-2}\cos\theta_{n-1},\\
            & ~\sin\theta_1\cdots\sin\theta_{n-2}\sin\theta_{n-1})~,
   \end{array}$$
 $$
 0\leq\theta_1,~\theta_2,~\cdots,~\theta_{n-2}\leq\pi~,~~~~~0\leq\theta_{n-1}
  \leq 2\pi~.
 $$
There are singularities in the transformation from $(x^0,{\bf x})$ to
$(\xi^0,{\bf x})$. Therefore, the spacetime structures of $AdS$ in the 
coordinate $(\xi^0,{\bf x})$ should be analysed carefully. The singularities 
occur at the submanifold  $x^0=0$.

The boundary $\sigma(x^i,x^j)=0$, besides the submanifold $x^0=0$,
is represented by $\xi^0\xi^0=\infty$ in the coordinate $(\xi^0,{\bf x})$. 
And the boundary of the submanifold $x^0=0$ is described by  $1-{\bf x}
{\bf x}^{\prime}=0$ in this coordinate. Then, we can describe the boundary 
of $AdS$ in the coordinate $(\xi^0,~{\bf x})$ as
\begin{equation}
\overline{M}^n:~~~~~\{\xi^0\xi^0=\infty,~1-{\bf xx}^{\prime}<0\}~\bigcup~
\{\xi^0\in(-\infty,~\infty),~1-{\bf xx}^{\prime}=0\}~.
\end{equation}

\begin{figure}
\epsfxsize=90mm
\epsfysize=90mm
\centerline{\epsffile{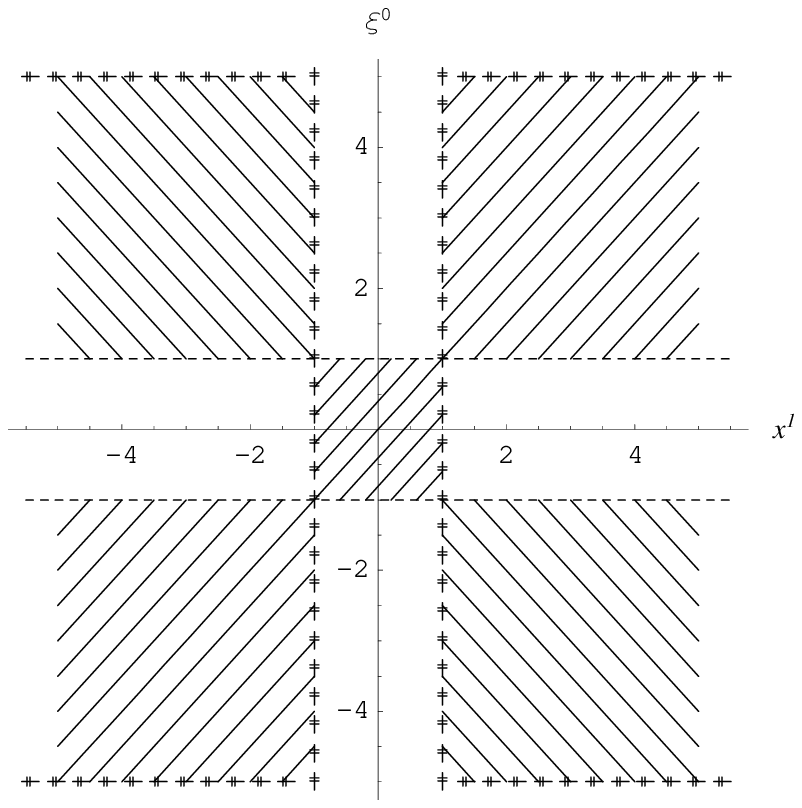}}
\vspace*{0.3cm}
\centerline{Fig.2~~$AdS_2$ in the coordinate $(\xi^0,~{\bf x})$ with boundary
                   denoted by dagger lines. }
\end{figure}

In figure 2, we give a plot of $AdS$ in the coordinate $(\xi^0,{\bf x})$.\\
By substituting the coordinate variables $\xi^0$ and $\rho$ with $\eta$ and
$\theta$,

\begin{equation}
\left\{
  \begin{array}{l}
  \rho=\tanh^{-1}\left[\frac{1}{2}\left(\tan(-\eta+\theta)-\tan(\eta+\theta-
        \pi)\right)\right]~, ~~~\frac{\pi}{2}>-\eta+\theta>\eta+\theta-\pi>
        -\frac{\pi}{2}  \\[0.5cm]
  \xi^0=\tanh^{-1}\left[\frac{1}{2}\left(\tan(-\eta+\theta)+\tan(\eta+\theta-
        \pi)\right)\right]~,~~~{\rm for}~1-\xi^0\xi^0<0,
         ~~1-\rho^2<0~;\\[0.7cm]
  \rho=\tanh\left[\frac{1}{2}\left(\tan(\eta+\theta)-\tan(\eta-\theta)\right)
       \right]~,  ~~~\frac{\pi}{2}>\eta+\theta>\eta-\theta>-\frac{\pi}{2} \\[0.5cm]
  \xi^0=\tanh\left[\frac{1}{2}\left(\tan(\eta+\theta)+\tan(\eta-\theta)\right)
        \right]~,~~~{\rm for}~1-\xi^0\xi^0>0,~~1-\rho^2>0,
\end{array}
\right.
\end{equation}
we can rewrite the metric (\ref{eq5}) into the form

\begin{equation}
\left\{
  \begin{array}{rcl}
  ds^2&=&\sinh^{-2}\left[\frac{1}{2}\left(\tan(\eta+\theta)-\tan(\eta-\theta)
        \right)\right]\\[0.5cm]
      & &\cdot\left[\cos^{-2}(\eta+\theta)\cos^{-2}(\eta-\theta)(d\eta^2-d
       \theta^2)-\cosh^2\left[\frac{1}{2}\left(\tan(\eta+\theta)+\tan(\eta-
       \theta)\right)\right]d{\bf u}d{\bf u}' \right]~,\\[0.4cm]
      & &~~~~~~~~~~~~~~~~~~~~~~~~~~~~~~~~{\rm for}~1-\xi^0\xi^0<0,
         ~~1-\rho^2<0~;\\[0.7cm]
  ds^2&=&\cosh^{-2}\left[\frac{1}{2}\left(\tan(\eta+\theta)+\tan(\eta-\theta)
       \right)\right]\\[0.5cm]
      & & \cdot\left[\cos^{-2}(\eta+\theta)\cos^{-2}(\eta-\theta)(d\eta^2-d
       \theta^2)-\sinh^2\left[\frac{1}{2}\left(\tan(\eta+\theta)-\tan(\eta-
       \theta)\right)\right]d{\bf u}d{\bf u}' \right]~,\\[0.5cm]
      & &~~~~~~~~~~~~~~~~~~~~~~~~~~~~~~~~{\rm for}~1-\xi^0\xi^0>0,
         ~~1-\rho^2>0~.
\end{array}\right.
\end{equation}
This form of metric allows us to draw the Penrose diagram\cite{011} of $AdS$.
The plot of the Penrose diagram of $AdS$ is presented in figure 3. A horizon
is shown clearly in the coordinate $(\xi^0,{\bf x})$.
In the coordinate $(\xi^0,{\bf x})$, $AdS$ is divided into two parts by the
horizon, {\em i.e.}, region I $\{\xi^0\xi^0<1,~{\bf xx}^{\prime}<1\}$ and
region II $\{\xi^0\xi^0>1,~{\bf xx}^{\prime}>1\}$. Different behaviors of
physical systems in the two regions should be investigated, separately.

\begin{figure}
\epsfxsize=90mm
\epsfysize=90mm
\centerline{\epsffile{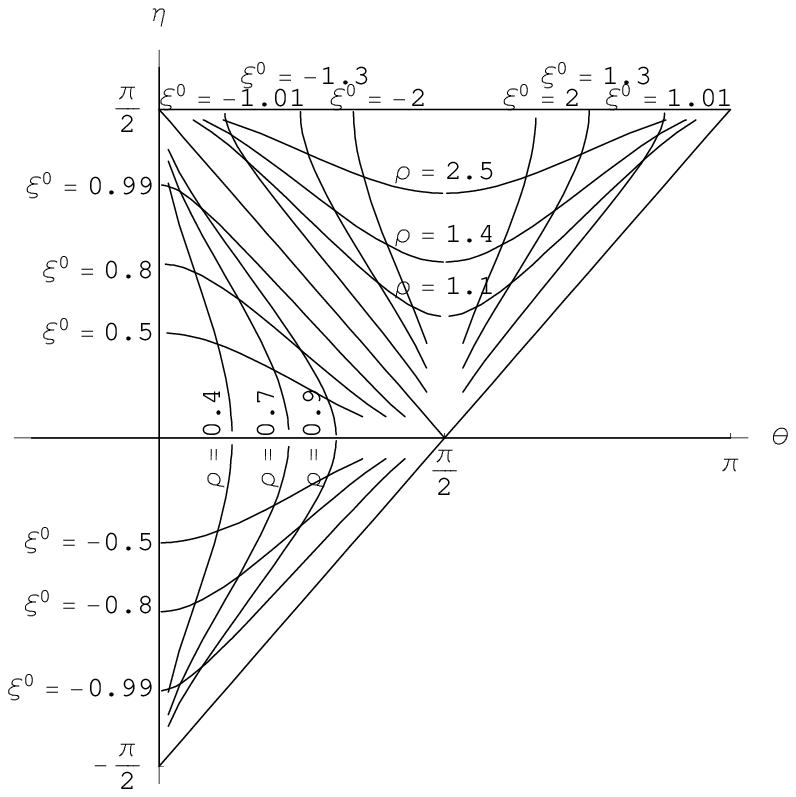}}
\vspace*{0.3cm}
\centerline{Fig.3~~The Penrose diagram of $AdS$.}
\end{figure}

\section{Exact solutions  in the region I}

$\quad$ In this section, we shall solve exactly the equation of motion for
massive scalar field in the region I.
In the region, the Laplace operator is of the form

\begin{eqnarray}
\Box&=&\left( 1-\xi^0\xi^0 \right)\partial_{\xi^0}^2-(n+1)\xi^0\partial_{\xi^0}
     \nonumber\\
  &&-\left( 1-\xi^0\xi^0 \right)^{-1}\left[ \left(1-\rho^2 \right)^2
     \partial^2_\rho+\rho^{-1}\left( 1-\rho^2 \right)\left(n-1-2\rho^2 \right)
     \partial_\rho \right] \\
  &&-\left( 1-\xi^0\xi^0 \right)^{-1}\left( 1-\rho^2 \right)\rho^{-2}
     \partial^2_{\bf u}~,\nonumber
\end{eqnarray}
where $\partial^2_{\bf u}$ denotes the angular part of the Laplace operator.

To solve the equation of motion of scalar field (Klein-Gordon equation)

 \begin{equation}
  \left(~\Box+m_0^2~\right)~\Phi(\xi^0,{\bf x})=0~,
 \end{equation}
we write the field variable $\Phi(\xi^0,{\bf x})$ into the form

 \[ \Phi(\xi^0,\rho,{\bf u})=T(\xi^0)U(\rho)Y({\bf u})~. \]
This form of field variable transforms the equation of motion into three 
separate parts\cite{11a},

 \begin{eqnarray}
  && \left[(1-\xi^0\xi^0)^2\partial^2_{\xi^0}-(n+1)\xi^0(1-\xi^0\xi^0)
     \partial_{\xi^0}+m_0^2(1-\xi^0\xi^0)+(\epsilon^2-m_0^2)\right]T(\xi^0)=0~,
     \nonumber\\[0.5cm]
  && \left[\partial^2_\rho+ \displaystyle\left[\frac{2}{\rho}+\frac{n-3}
     {\rho(1-\rho^2)}\right]\partial_\rho-\left[\frac{m_0^2-\epsilon^2}
     {(1-\rho^2)^2}+\frac{l(l+n-2)}{\rho^2(1-\rho^2)}\right]\right]U(\rho)=0~,
     \\[0.5cm]
  && \left[l(l+n-2)+\partial^2_{\bf u}\right]Y_{l{\bf m}}({\bf u})=0~,\nonumber
 \end{eqnarray}
where $Y_{l{\bf m}}({\bf u})$ is the spherical harmonic on the sphere
$S^{n-1}$~ and $\epsilon$ is a constant(physical meanings of it should be
discussed later). \\
For the radial field variable, we can write the solutions as the form
$$U(\rho)=\rho^l(1-\rho^2)^{\mu/2}{\cal F}(\rho^2)~,$$
with $\mu$ satisfies
$$\begin{array}{rl}
    &\mu^2-(n-1)\mu-(m_0^2-\epsilon^2)=0~,\\
    {\rm or}~~~~~~~&\\
    &\mu_{\pm}=\displaystyle \frac{n-1}{2}\left(1\pm\sqrt{1+\frac{4(m_0^2-
                   \epsilon^2)}{(n-1)^2}}\right)~.
   \end{array}
$$
and ${\cal F}$ satisfies

\begin{equation}
 y(1-y){\cal F}''(y)+\left[l+\frac{n}{2}-(l+\mu+\frac{3}{2})y\right]{\cal F}'(y)
 -\left[\frac{1}{4}l(l+1)+\frac{1}{4}\mu(\mu+1)+\frac{1}{2}l\mu\right]
 {\cal F}(y)=0~.
\end{equation}
This is a hypergeometrical equation. Then, we get the radial variable for 
the field 

 \begin{equation}
  U(\rho)=\rho^l(1-\rho^2)^{\mu/2}F\left(\frac{1}{2}(l+\mu+1),~\frac{1}{2}
                                   (l+\mu),~l+\frac{n}{2};~\rho^2\right)~.
 \end{equation}
For the time-like part of the field variable, let
$$T(\xi^0)\equiv(1-\xi^0\xi^0)^{(1-n)/4}P(\xi^0)~,$$
we have an associated Legendre equation
$$
 (1-\xi^0\xi^0)P''(\xi^0)-2\xi^0P'(\xi^0)+\left[-\frac{1}{4}(1-n^2)+
 m_0^2-\displaystyle\frac{\frac{1}{4}(n-1)^2-
 (\epsilon^2-m_0^2)}{1-\xi^0\xi^0}\right]P(\xi^0)=0~.
$$
Solutions of time-like part of the field variable, thus, are of the forms

 \begin{equation}
  T(\xi^0)=\left(1-\xi^0\xi^0\right)^{(1-n)/4}
            \cdot\left\{
            \begin{array}{l}
             P_{\nu}^{\nu^{\prime}}(\xi^0)~,\\
                            \\
             Q_{\nu}^{\nu^{\prime}}(\xi^0) ~,
            \end{array}
      \right.
 \end{equation}
where $\nu,~\nu^{\prime}~$satisfy
 \begin{equation}
 \begin{array}{l}
  \nu(\nu+1)=m_0^2-\displaystyle \frac{1}{4}(1-n^2)~,\\
  (\nu^{\prime})^2=\displaystyle \frac{1}{4}(n-1)^2-(\epsilon^2-m_0^2) ~.
 \end{array}
 \end{equation}
If $\nu^{\prime}$ and $\nu$ are integers, $T^{\nu^{\prime}}_\nu(\xi^0)$ is 
finite at the points $\xi^0=\pm 1$. This so-called natural boundary condition
gives rise to the discretization of the mass and energy spectrum of free 
scalar particles in this slice of $AdS$\cite{11a}--\cite{014}.

It should be noticed that as $\rho\rightarrow 1$~(approaches to the boundary),
the field varible $\Phi(x)$ has a singularity of the form
$(1-\rho^2)^{\mu_-/2}$~.


\section{Bulk-boundary propagator in the region I}

$\quad$ In the region I, we have

\begin{equation}
\begin{array}{r}
AdS_{n+1}\cap{\cal U}_1:~~~~~1-{\bf x}{\bf x}^{\prime}>0~,\\[0.5cm]
\overline{M}^n\cap{\cal U}_1:~~~~~1-{\bf x}{\bf x}^{\prime}=0~,
\end{array}
\end{equation}
or, equivalently,

$$AdS_{n+1}=RP^1\times B^n~.$$
As shown in the previous section, the time-like coordinate ($\xi^0$) and 
space-like coordinates (${\bf x}$) part of solutions of the motion can 
be separated and we have 

\begin{equation}\label{equation}
\begin{array}{rcl}
(\Box+m_0^2)\Phi(x)&=&0\\[0.5cm]
                 &=&(\Delta-(m_0^2-\epsilon^2))T(\xi^0)\Phi({\bf x})~.
\end{array}
\end{equation}
Here $\Delta$ is the Laplace operator in the $n$-dimensional unit ball
$B^n$ with the Hua's metric,

\begin{equation}
\Delta=\frac{1}{\sqrt{-g}}\sum_{\alpha,\beta=1}^n\frac{\partial}
{\partial x^\alpha}\left(
\sqrt{-g}g^{\alpha\beta}
       \frac{\partial}{\partial x^\beta}\right)~.
 \end{equation}
This fact tells us that the equal `time' surface of  $AdS_{n+1}$ is a
unit ball with  conformal flat or Hua's metric\cite{015}.
Equation(\ref{equation}) shows that the equations of motion for scalar
fields with mass $m_0$ in
AdS$_{n+1}$ reduces to the equations of motion for scalar fields with mass
$\sqrt{m_0^2-\epsilon^2}$ on the equal time surface: $B^n$. At the case of
$\sqrt{m_0^2-\epsilon^2}=0$, the bulk-boundary propagator can be obtained 
by a standard method\cite{016}

\begin{equation}
G^E_{B\partial}({\bf x},{\bf u})=\frac{(1-{\bf x}{\bf x}^{\prime})^
    {\frac{n-1}{2}}}{(1-{\bf u}{\bf x}^{\prime})^{n-1}}~.
\end{equation}
The bulk-boundary propagator for massive scalar fields is of the form

\begin{equation}
G^\pm_{B\partial}({\bf x},{\bf u})=\left(G^E_{B\partial}({\bf x},{\bf u})
\right)^{\alpha_\pm}~,
\end{equation}
where we have used the notation
$$
\alpha_\pm=\displaystyle\frac{1}{2}\left(1\pm\sqrt{1+\frac{4(m_0^2-\epsilon^2)}
           {(n-1)^2}}\right)~.
$$
The bulk field $\Phi(x)$ determined by the bulk-boundary propagator
$G^\pm_{B\partial}({\bf x},{\bf u})$ and the field living on the boundary
$\phi({\bf u})$ is of the form 

\begin{equation}\label{formula1}
\Phi^\pm(x)=\frac{1}{\omega_{n-1}}\int_{{\bf uu}^{\prime}=1}\cdots\int d{\bf u}
G^\pm_{B\partial}({\bf x},{\bf u})T(\xi^0)\phi({\bf u})~.
\end{equation}\\
As a $\delta$-function, $G_{B\partial}^E({\bf x},{\bf u})$ is divergent with 
dimension $1-n$, so the divergent dimension of $G_{B\partial}^{\pm}({\bf x},
{\bf u})$ is $(1-n)\alpha_{\pm}$, and the bulk field $\Phi^{+}(x)$ determined 
this way is divergent at the boundary with dimension

\begin{equation}
d=(1-n)(\alpha_{+}-1)=\frac{n-1}{2}\left(1-\sqrt{1+\frac{4(m_0^2-\epsilon^2)}
{(n-1)^2}}\right)~.
\end{equation}
It should be noticed that the behavior of the bulk field $\Phi^{+}(x)$ at the
boundary is the same as that of the exact results for the Klein-Gordon
equation. At this sense, there is a one-to-one correspondence between the
fields got by constructing the bulk-boundary propagator and ones obtained by 
solving exactly the equation of motion. This indicates that the dynamics for 
massive scalar fields can be probed by making use of the $AdS/CFT$ 
correspondence.  And at the same foot, the conformal field theory living on 
the boundary can be investigated by dynamics in the bulk by making use of 
the $AdS/CFT$ correspondence.

\section{Exact solutions in the region II}

$\quad$ To solve the equation of motion for massive scalar field in the 
region II, it is convenient to introduce the variables $\varsigma^0$ and 
$r$ as
$$
\varsigma^0\equiv\frac{1}{\xi^0}~,~~~~~r\equiv\frac{1}{\rho}~.
$$
In terms of $\varsigma^0$, $r$ and ${\bf u}$, the $SO(n,2)$ invariant metric
(\ref{eq5}) can be rewritten as

\begin{equation}
 ds^2=\frac{1}{\varsigma^0\varsigma^0}\left[ \frac{d\varsigma^0d\varsigma^0
 }{\varsigma^0
 \varsigma^0-1}-\left( \varsigma^0\varsigma^0-1 \right)
      \left[\frac{dr^2}{\left(r^2-1\right)^{2}}+ \frac{d{\bf u}d{\bf u'}}{r^2-1}
      \right]\right]~.
\end{equation}
The corresponding Laplace operator is

\begin{eqnarray}
\Box&=&\varsigma^0\varsigma^0(\varsigma^0\varsigma^0-1)
\partial^2_{\varsigma^0}+[2\varsigma^0\varsigma^0\varsigma^0+(n-1)\varsigma^0]
\partial_{\varsigma^0}\nonumber\\
       &&\displaystyle +\frac{\varsigma^0\varsigma^0}{1-\varsigma^0\varsigma^0}
         \left[(r^2-1)^2\partial^2_r+(3-n)r(r^2-1)\partial_r \right]\\
       &&\displaystyle +\frac{\varsigma^0\varsigma^0}{1-\varsigma^0
       \varsigma^0}(r^2-1)\partial^2_{\bf u}~.\nonumber
\end{eqnarray}
To solve the equation of motion in the region II

\begin{equation}
\left(~\Box+m^2_0~\right)\Phi(\varsigma^0,r,{\bf u})=0~,
\end{equation}
we write the scalar field variable into the form

$$
\Phi(\varsigma^0,r,{\bf u})=T(\varsigma^0)R(r)Y({\bf u})~.
$$
And the equation of motion is transformed as

\begin{equation}
\begin{array}{l}
\left[(1-r^2)^2\partial_r^2+(n-3)r(1-r^2)\partial_r+l(l+n-2)(1-r^2)
+(\epsilon^2-m_0^2)\right]R(r)=0~,\\[0.5cm]
\left[(\varsigma^0\varsigma^0-1)^2\partial_{\varsigma^0}^2+\left(2\varsigma^0
+\displaystyle\frac{n-1}{\varsigma^0}\right)(\varsigma^0\varsigma^0-1)\partial
_{\varsigma^0}+m_0^2\displaystyle\frac{\varsigma^0\varsigma^0-1}{\varsigma^0
\varsigma^0}+(\epsilon^2-m_0^2)\right]T(\varsigma^0)=0~,\\[0.5cm]
\left[l(l+n-2)+\partial^2_{\bf u}\right]Y_{l{\bf m}}({\bf u})=0~.
\end{array}
\end{equation}
Solutions, in terms of the hypergeometrical functions, are of the forms

\begin{equation}
\begin{array}{l}
 \displaystyle T(\xi^0)=(\xi^0\xi^0)^{-(\mu+\nu)/2}
 (\xi^0\xi^0-1)^{\nu/2}
 F\left(\frac{1}{2}(\mu+\nu)~,\frac{1}{2}(\mu+\nu+1)~,
          \mu-\frac{n}{2}+1~;(\xi^0\xi^0)^{-1}\right)~,\\
 \displaystyle R(\rho)  =\rho^{-\lambda}(\rho^2-1)^{\lambda/2}F\left(\frac{1}{2}(\lambda-n-l)+1~,\frac{1}{2}
          (\lambda+l)~,\frac{1}{2}~;\rho^{-2}\right)~,
\end{array}
\end{equation}
where the notations $\mu,~\nu$ and $\lambda$ have been used
\begin{equation}\label{notation}
\begin{array}{l}
 \displaystyle \mu_{\pm}=\frac{n}{2}\left(1\pm \sqrt{1+\frac{4m_0^2}{n^2}}\right)  ~,\\
 \displaystyle \nu_{\pm}=\frac{n-1}{2}\left(-1\pm\sqrt{1+
 \frac{4(m_0^2-\epsilon^2)}{(n-1)^2}}\right)  ~,\\
 \displaystyle\lambda_{\pm}=\frac{n-1}{2}\left(1\pm \sqrt{1+
 \frac{4(m_0^2-\epsilon^2)}{(n-1)^2}}\right)  ~.
\end{array}
\end{equation}
We notice that  the solutions have a singularity of the form
$(\xi^0\xi^0)^{-\mu_-/2}$ at the boundary  ($\xi^0=\infty$).
For the other part of the boundary ($\rho=1$), these solutions have a 
singularity of the form $(\rho^2-1)^{\lambda_-/2}$. This is in agreement 
with behaviors, on the boundary, of the exact solutions and the fields 
obtained by making use of the bulk-boundary propagator in the region I.


\section{Bulk-boundary propagator in region II}

$\quad$ One of the keystones of the $AdS/CFT$ duality is the bulk-boundary 
propagator. It should be noticed that the bulk-boundary propagator for $AdS$
can not be obtained directly by a Wick rotation from what got for the Euclidean
version of $AdS$, singularities appear in the bulk and indeed do not satisfy
neccessary conditions for a propagator. To discuss the $AdS/CFT$ correspondence
seriously, we have to construct a bulk-boundary propagator for $AdS$ but not 
for the Euclidean version of $AdS$. In fact, we have presented a bulk-boundary
propagator for the region I in section $5$. Here, we write down a bulk-boundary
propagator for the region II,\\ \\
$G_{B\partial}^\pm({\bf u},\rho,\xi^0;{\bf v},\varrho)$
\begin{equation}
\begin{array}{r}
=\displaystyle\int d\epsilon\sum_l\sum_{\bf m}Y_{l{\bf m}}({\bf u}-{\bf v})
(\xi^0\xi^0)^{-(\mu+\nu)/2} (\xi^0\xi^0-1)^{\nu/2}
(\rho-\varrho)^{-\lambda}((\rho-\varrho)^2-1)^{\lambda/2}\\
\times F\left(\frac{\mu+\nu}{2},\frac{\mu+\nu+1}{2},
          \mu-\frac{n}{2}+1~;(\xi^0\xi^0)^{-1}\right)
 F\left(\frac{\lambda-n-l}{2}+1,\frac{\lambda+l}{2},\frac{1}{2}~;
 (\rho-\varrho)^{-2}\right)~,
\end{array}
\end{equation}
where $\mu$, $\nu$ and $\lambda$ take the same values as in the equation
(\ref{notation}).
It is not difficult to show that this bulk-boundary prpagator satisfies the
equation of motion for a scalar field with mass $m_0$,

\begin{equation}
\left(~\Box+m^2_0~\right)G_{B\partial}^\pm({\bf u},\rho,\xi^0;{\bf v},\varrho)=0~,
\end{equation}
in the region II of $AdS$. The bulk field can be determined from a field 
living on the boundary  by making use of the bulk-boundary propagator
$G^\pm_{B\partial}({\bf u},\rho,\xi^0;{\bf v},\varrho)$. To
display this fact explicitly, we notice that the boundary of $AdS_{n+1}$ is
the compactification of the $n$-dimensional Minkowski space $\overline{M}_n$.
Then, at the boundary of $AdS$, we can introduce a metric \cite{017} conformal
to the Lorentzian metric, so that, the integration on the boundary of $AdS$ 
can get a finite result,

\begin{equation}
ds^2=\frac{\displaystyle\sum_{i,j=0}^{n-1}\eta_{ij}d\chi^id\chi^j}
{\left[1+\left(\chi^0-\displaystyle\sqrt{\sum_{i=1}^{n-1}\chi^i\chi^i}\right)^2\right]
 \left[1+\left(\chi^0+\displaystyle\sqrt{\sum_{i=1}^{n-1}\chi^i\chi^i}
 \right)^2\right]}~.
\end{equation}
The total volume of the boundary of $AdS$ is finite respect to this metric

\begin{equation}
\begin{array}{rcl}
\omega_n&\equiv&\displaystyle\int_{{\bf vv'}=1}\cdots\int d{\bf v}\int d\varrho\\[1cm]
        &=&\displaystyle\int_{\overline{M}_n}
           \frac{d\chi^0\wedge d\chi^1\wedge\cdots\wedge d\chi^{n-1}}
           {\left[1+\left(\chi^0-\displaystyle\sqrt{\sum_{i=1}^{n-1}\chi^i
           \chi^i}\right)^2\right]^{n/2}
 \left[1+\left(\chi^0+\displaystyle\sqrt{\sum_{i=1}^{n-1}\chi^i\chi^i}
 \right)^2\right]^{n/2}}~.
 \end{array}
\end{equation}
Then, we can say that there is a totally determined bulk field $\Phi^\pm(x)$
corresponding to each field which living on the boundary by making use of the
bulk-boundary propagator,

\begin{equation}\label{formula}
\Phi^\pm(x)=\frac{1}{\omega_{n}}\int_{\bf vv'=1}\cdots\int d{\bf v}\int d\varrho
G^\pm_{B\partial}({\bf u},\rho,\xi^0;{\bf v},\varrho)\phi({\bf v},\varrho)~.
\end{equation}
This shows that $G^\pm_{B\partial}({\bf u},\rho,\xi^0;{\bf v},\varrho)$ is 
really a bulk-boundary propagator in $AdS$ for massive scalar field.  The bulk 
field $\Phi(x)$ got from this way does not approach to $\phi({\bf v},\varrho)$
when limited on the boundary, but goes divergent at boundary with dimension

\begin{equation}
d=\frac{n}{2}\left(1-\sqrt{1+\frac{4m_0^2}{n^2}}\right)~.
\end{equation}
This is in good agreement with the exact solutions got in previous section.

\vskip 2mm

\centerline{\bf Acknowledgments}

Two of us (Z.C. and H.Y.G.) would like to thank Y. H. Gao,
C. G. Huang, S. K. Wang, K. Wu  and C. J. Zhu for
enlightening discussions.
The work was supported in part by the National Natural Science Foundation of China.


\begin{thebibliography}{999}
\bibitem{001} See, for example, P. A. M. Dirac, ``The electronic wave equation
              in de-Sitter space'', Ann. Math. {\bf 36} (1935) 657;\\
              C. Fronsdal, ``Elementary particles in a curved space'', Rev.
              Mod. Phys. {\bf 37} (1965) 221;\\
              Q. K. Lu, Z. L. Zou and H. Y. Guo, ``The kinematic effects in the classical domains and the red-shift phenomena of extra-galactic objects",
              Acta Physica Sinica (In Chinese) {\bf 23} (1974) 225.
\bibitem{002} J. Maldacena, ``The large $N$ limit of superconformal field
              theories and supergravity'', Adv. Theor. Math. Phys. {\bf 2}
              (1998) 231, hep-th/9711200.
\bibitem{003} S. S. Gubser, I. R. Klebanov and A. M. Polyakov, ``Gauge theory
              correlators from non-critical string theory'', Phys. Lett.
              B{\bf 428} (1998) 105, hep-th/9802109.
\bibitem{004} E. Witten, ``Anti de Sitter space and holography'', Adv. Theor.
              Math. Phys. {\bf 2} (1998) 253, hep-th/9802150.
\bibitem{03a} A. M. Polyakov, ``String theory and quark confinement'', Nucl.
              Phys. Proc. Suppl. {\bf 68} (1988) 1, hep-th/9711002.
\bibitem{04a} A. M. Polyakov, ``The wall of the cave'', Int. J. Mod. Phys.
              A{\bf 14} (1999) 645, hep-th/9809057.
\bibitem{005} E. Witten, ``Anti-de Sitter space, thermal phase transition, and
              confinement in gauge theories'', Adv. Theor. Math. Phys. {\bf 2}
              (1998) 505, hep-th/9803131.
\bibitem{006} C. Csaki, H. Ooguri, Y. Oz, J. Terning, ``Glueball mass spectrum
              from supergravity'', JHEP {\bf 9901} (1999) 017, hep-th/9806021.
\bibitem{007} R. M. Koch, A. Jevicki, M. Mihailescu, J. P. Nunes, ``Evaluation
              of glueball masses from supergravity'', Phys. Rev. D{\bf 58}
              (1998) 105009, hep-th/9806125.
\bibitem{008} M. Zyskin, ``A note on the glueball mass spectrum'', Phys. Lett.
              B{\bf 439} (1998) 373, hep-th/9806128.
\bibitem{009} J. A. Minahan, ``Glueball mass spectra and other issues for
              supergravity duals of QCD models'', JHEP {\bf 9901} (1999) 020,
              hep-th/9811156.
\bibitem{010} For a review, see, O. Aharony, S.S. Gubser, J. Maldacena, H. Ooguri
              and Y. Oz, ``Large $N$ field theories, string theory and
              gravity'', Physics Reports {\bf 323} (2000) 183, hep-th/9905111.
\bibitem{10a} L. K. Hua, ``Harmonic analysis of functions of several
              complex variables in the classical domains'', Trans. Am. Math.
              Soc. {\bf 6} (1963).
\bibitem{10b} Q. K. Lu, ``Classical manifolds and classical domains'' (in
              Chinese), Sci.\& Tech. Press (Shanghai) (1963).
\bibitem{10c} H.Y. Kuo (H.Y. Guo), `` The groups of transformations and
              invariants for the typical spacetime'', Kexue Tongbao (Comm. in
              Science) (In Chinese) {\bf 22} (1977) 487.
\bibitem{011} B. Carter, ``Complete analytic extension of the symmetry axis
              of Kerr's solution of Einstein's equation'', Phys. Rev.
              {\bf 141} (1963) 1242.
\bibitem{11a} G. Y. Li and H. Y. Guo, ``Relativistic quantum mechanics for
              scalar particles and spinor particles in the Beltrami-de Sitter
              spacetime'' (in Chinese), Acta Physica Sinica
             {\bf 31} (1982) 1501.
\bibitem{11b} P. Breitenlohner and D. Z. Freedman, ``Positive energy in Anti-de
              Sitter backgrounds and gauged extended supergravity'', Phys.
              Lett. {\bf 115}B (1982) 197.
\bibitem{012} P. Breitenlohner and D. Z. Freedman, ``Stability gauged extended
              supergravity'', Ann. Phys. {\bf 144} (1982) 249.
\bibitem{013} V. Balasubramanian, P. Kraus and A. Lawrence, ``Bulk vs. boundary
              dynamics in Anti-de Sitter space-time'', Phys. Rev. D{\bf 59}
              (1999) 046003, hep-th/9805171.
\bibitem{014} V. Balasubramanian, S. B. Giddings and A. Lawrence, ``What do
              CFTs tell us about Anti-de Sitter spacetimes?'', JHEP {\bf 9903}
              (1999) 001, hep-th/9902052.
\bibitem{015} Q. K. Lu, Z. Chang and H. Y. Guo, ``Global geometric properties
              of AdS and the AdS/CFT correspondence'', hep-th/0008228.
\bibitem{016} Z. Chang and H. Y. Guo, ``Symmetry realization, Poisson kernel
              and the AdS/CFT correspondence'', Mod. Phys. Lett. A{\bf 15}
              (2000) 407, hep-th/9910136.
\bibitem{017} Q. K. Lu, ``The Yang-Mills fields on the Minkowski space'',
              Science in  China (Series A), Vol. {\bf 41} (1998) 1061.

\end{thebibliography}
\end{document}